\documentclass[preprint]{imsart}

\usepackage{amsthm,amsmath,graphicx}

\startlocaldefs
\endlocaldefs

\begin{document}

\begin{frontmatter}

\title{Risk models for breast cancer and their validation}
\runtitle{Breast cancer risk models}

    \author{\fnms{Adam} R. \snm{Brentnall}\ead[label=e1]{a.brentnall@qmul.ac.uk}},
    \author{\fnms{Jack} \snm{Cuzick}\corref{}\ead[label=e2]{j.cuzick@qmul.ac.uk}}

    \address{Centre for Cancer Prevention, Wolfson Institute of Preventive Medicine, Queen Mary University of London, Charerhouse square, London, EC1M 6BQ \printead{e1,e2}.}

    \runauthor{Brentnall \& Cuzick}

\begin{abstract}
     Strategies to prevent cancer and diagnose it early when it is most treatable are needed to reduce the public health burden from rising disease incidence. Risk assessment is playing an increasingly important role in targeting individuals in need of such interventions. For breast cancer many individual risk factors have been well understood for a long time, but the development of a fully comprehensive risk model has not been straightforward, in part because there have been limited data where joint effects of an extensive set of risk factors may be estimated with precision. In this article we first review the approach taken to develop the IBIS (Tyrer-Cuzick) model, and describe recent updates. We then review and develop methods to assess calibration of models such as this one, where the risk of disease allowing for competing mortality over a long follow-up time or lifetime is estimated. The breast cancer risk model model and calibration assessment methods are demonstrated using a cohort of 132 139 women attending mammography screening in Washington, USA.
   
\end{abstract}



\end{frontmatter}

\section{Introduction}

Unlike lung cancer (tobacco smoking) and cervix cancer (persistent infection with the human papilloma virus) where a single factor explains the majority of the cases, a large number of factors have been found to be important for determining the risk of breast cancer. An increased risk in women with a family history appears to have been known in ancient Roman times \cite{1991-steel}, and in 1842 Rigoni-Stern reported that nuns had an increased risk of breast cancer \cite{1842-RigoniStern,1987-destavola}, paving the way for further research which established that not having a child or having a first childbirth at an older age increased the risk of this disease. Much of the modern work on risk factors was done by Macmahon and colleagues, mostly at Harvard \cite{1969-macmahon}. Pike and colleagues emphasized the link with oestrogens and proposed that much of the population risk could explained by a computing the cumulative lifetime exposure to oestrogens based mostly on age at menarche, childbearing history and age at menopause \cite{1983-pike}; an important meta-analysis and update on these and other factors was provided by Beral and colleagues \cite{2012_collabovmenop}. 

In 1976 Wolfe discovered that breast density was another key risk factor that was broadly unrelated to the known classical factors, but of roughly equal importance in their combined predictive value \cite{1976_Wolfe}. This has been further developed by Boyd \cite{1982-boyd,2007_Boydetal}, McCormack and others \cite{2006_McCormackSilva}. In the same paper Wolfe also suggested \cite{1976_Wolfe} using breast density for risk assessment to determine how best to use mammography for breast cancer screening.  Several decades later, advances in the ability to accurately stratify women into higher and lower-risk groups are likely to move early detection strategies towards this vision, and replace `one-size-fits-all' screening with so-called `risk-adapted' programs in which both the frequency with which a woman is screened and the modality are chosen based on the risk of breast cancer. Motivations include the identification of women at extremely high risk, who are potential candidates for risk-reducing surgery or preventive therapy \cite{2015-eastonpharoah}, delineation of a group at moderately enhanced risk who might benefit from enhanced screening \cite{2010_FH01}, and identification of populations at sufficiently low risk so as to require less frequency or even no screening. 

A number of risk models have been developed, mostly for Caucasian women living in North America and Western Europe \cite{1989_Gailetal,1993-claus,2008_Antoniouetal}. In this article we focus on risk assessment using the Tyrer-Cuzick model \cite{2004_TyrerCuzick}. This is a hybrid of two popular sub-models often used for breast cancer risk assessment: a genetic segregation model for familial risk that is combined with a proportional-hazards regression model for other risk factors. 

Breast cancer risk factors in the model broadly fit into five general categories: (1) family history and highly penetrant dominant genetic mutations; (2) factors associated with oestrogen exposure, including age at first childbirth, age at menopause and menarche (beginning of periods), use of hormone replacement therapy, height and weight; (3) certain types of prior benign breast disease,  (4) breast imaging features seen on the mammogram - notably the amount of dense tissue (opaque areas on a mammogram or breast x-ray); and (5) common but individually less penetrant genetic differences (single nucleotide polymorphisms, SNPs), where several hundred relatively common genetic variants that individually have a small impact on disease have been identified, and that jointly make an important contribution to overall risk assessment  via a `polygenic risk score' \cite{2018-snp313}. There are also some other apparent risk factors which are harder to quantify, but which may improve the performance of the model. 

In the rest of this paper we outline the statistical basis of our model, review and develop methods to assess model calibration that include competing mortality (which has been handled differently in different papers), and apply our model and calibration assessment methods to data from a large cohort of women attending mammographic screening in Washington, USA.

\section{Breast cancer risk assessment}

\subsection{Background}

To provide an assessment of the risk of disease occurring within a person's residual lifetime it is important to take into account competing risks that could lead to death from other causes. In a classical `latent lifetimes' framework \cite{2001_Crowder} each individual is subject to $m$ potential causes of death with times $T_j$ ($j=1,\ldots,m$) and we observe their actual time of death $\tilde{T}=\textrm{min}(T_1, \ldots, T_m)$ and the specific cause $J=1,\ldots,m$. This framework assumes that each death is attributable to a single cause, or a defined group of causes. We extend this approach to also include the incidence of breast cancer, which is our principle interest, and do not consider death after breast cancer occurs. 

Multiple modes of failure are conveniently characterised using functions for the rate at which each cause $J$ occurs at each follow-up time, given that the person has not yet died, or experienced a specific event of interest. More precisely, this \textit{cause-specific hazard} for $j=1,\ldots,m$ is defined as 
\begin{eqnarray}
    h_j(t) & = & \lim_{\Delta \to 0} \frac{\textrm{P}(T_j < t + \Delta \mid \tilde{T} \geq t)}{\Delta}.\label{eqn:crudehaz}
\end{eqnarray}
Equation (\ref{eqn:crudehaz}) is estimable due to conditioning on $(\tilde{T} \geq t)$. Another measure of risk that is sometimes considered is the proportion of the population who have event $J$ when followed up to time $t$. This \textit{cumulative incidence function} is defined by  
\begin{eqnarray}
    \textrm{P}(J=j, \tilde{T} \leq t) & = & \int_0^t h_j(u) S(u) \textrm{d}u,\label{eqn-sub}
\end{eqnarray}
where 
\begin{eqnarray*}
    S(t) & = & \textrm{P}(\tilde{T} > t) \\ & = & \textrm{exp}\left\{ - \int_0^t \sum_{k=1}^{m} h_k(u) \textrm{d}u \right\}
\end{eqnarray*}
is the marginal (all-cause) survivor function for $\tilde{T}$. When there is just one cause (death, say) then (\ref{eqn-sub}) is the cumulative distribution function (for time of death). 

Breast cancer risk models within this framework typically consider two causes ($m=2$): $T_1$, the time to diagnosis of breast cancer and $T_2$, the time to death from other causes (excluding breast cancer mortality, which can't occur before the time of diagnosis), so that $\tilde{T} = \textrm{min}(T_1, T_2)$. Risk assessment for a women usually takes place when current age $t_0>20$ years, and the aim is to assess the absolute risk of breast cancer between age $t_0$ to age $t$, conditional on $q$ risk factors $\boldsymbol{x} = (x_1,\ldots,x_q)$. This may be done by extending (\ref{eqn-sub}) to condition on risk factors $\boldsymbol{x}$ and age at risk assessment $t_0$ through: 
\begin{eqnarray}
\textrm{P}(J=1, \tilde{T} \leq t \mid  \boldsymbol{x}, \tilde{T} > t_0 ) & = & \int_{t_0}^{t} h_1(u \mid \boldsymbol{x}) \textrm{exp}\left[ - \int_{t_0}^{u} \{h_1(v \mid \boldsymbol{x}) + h_2(v \mid \boldsymbol{x})\} \textrm{d}v \right]\textrm{d}u \nonumber \\ & \equiv & P_x(t_0,t), \label{eqn:absrisk} \end{eqnarray}
where $h_1(t \mid  \boldsymbol{x})$ is the conditional hazard of breast cancer at age $t$ (\textit{c.f.} equation \ref{eqn:crudehaz}) and $h_2(t \mid \boldsymbol{x})$ is the conditional hazard for competing mortality. In practice, and following \cite{1989_Gailetal}, competing mortality has often been taken to depend only on age and has been calibrated using national mortality statistics excluding breast cancer incidence. This is reasonable for breast cancer, as it is not strongly linked other causes of death, but is more problematic, \textit{e.g.} for lung cancer, where tobacco smoking is a major factor for it and other causes of death such as cardiovascular disease.  

Breast-cancer specific hazards have been estimated in two ways:
 \begin{enumerate}
     \item  Regression models. Risk is frequently calculated by combination of a regression model derived from case-control or cohort studies of specific relative risks combined with absolute population-based incidence rates from cancer registries. An early and widely used example of this is the Gail model \cite{1989_Gailetal}, which has been further developed in the Breast Cancer Surveillance Consortium (BCSC) model \cite{2006_Chenetal, 2008_Ticeetal,2010_Mealiffeetal}.
     \item Estimation from family pedigree data of the probability of carrying one or more high-risk mutations using segregation analysis, and then using the penetrance of a mutation to alter age-specific risk \cite{1993-claus}. Two examples are BRCAPRO and BOADICEA \cite{1998-parmigiani,2008_Antoniouetal}.
                   \end{enumerate}
A novel aspect of the Tyrer-Cuzick model is that it employs both methods \cite{2004_TyrerCuzick}. The segregation model estimates the hazard function $h_G(t\mid \boldsymbol{x}_1)$ at age $t$ due to genetic factors, conditional on information (denoted $\boldsymbol{x}_1$) about a woman's family history (family tree) of breast and/or ovarian cancer, and results of any tests in the family from known highly penetrant breast cancer genes such as \textit{BRCA1} and \textit{BRCA2}. This is then combined with a relative hazard regression function $r(\boldsymbol{x}_2)>0$ based on other risk factors ($\boldsymbol{x}_2$) through 
\begin{eqnarray}
h_1(t\mid \boldsymbol{x}_1, \boldsymbol{x}_2) & = & h_G(t\mid \boldsymbol{x}_1)r(\boldsymbol{x}_2). \label{eqn_phmod1}
\end{eqnarray}
We divide family history into two factors $\boldsymbol{x}_1 = \{ \boldsymbol{x}_1^{(a)},\boldsymbol{x}_1^{(b)}\}$ corresponding to (a) breast and (b) ovarian cancer. The vector $\boldsymbol{x}_1^{(.)}$ has components for event time, censoring information and family relationships, and $\boldsymbol{x}_2$ is based on the other risk factors shown in Table 1.
Risk models that only account for family history based on a segregation model do not use $\boldsymbol{x}_2$. Regression-only models developed are often proportional-hazards models of the same form as (\ref{eqn_phmod1}), but with age-specific population rates instead of $h_G(.)$; partial information on family history of breast cancer can be included in the regression function, \textit{e.g.} by including a covariate for the number of affected first-degree relatives. 

\subsection{Tyrer-Cuzick segregation model}

\subsubsection{Statistical model}

The genetic risk breast cancer is modelled through estimates of mutations in \textit{BRCA1} and \textit{BRCA2} genes and an unknown dominant gene. Some of the genes which make up the unknown component have been discovered \cite{2015-eastonetal}, but most have not, and are only inferred. Even the known ones are rare and individually do not make a major contribution, and are not so commonly tested for even in women with a family history. We use the variable $c_1$ for \textit{BRCA1/2} status ($0$ if not a \textit{BRCA1/2} carrier, $1$ if a \textit{BRCA1} carrier, and $2$ if \textit{BRCA2} carrier; joint carriers are extremely rare and are modelled as \textit{BRCA1} carriers). The unknown dominant gene is denoted $c_2$ ($0,1$ if respectively not a carrier or a carrier). The model makes the following assumptions. 
\begin{enumerate}
    \item[A.i] Breast $\boldsymbol{x}_1^{(a)}$ and  ovarian cancer $\boldsymbol{x}_1^{(b)}$ family histories are conditionally independent given $c_1$ and $c_2$:
    \begin{eqnarray*}
p(\boldsymbol{x}_1\mid c_1,c_2) & = & p\{\boldsymbol{x}_1^{(a)}\mid c_1,c_2\}p\{\boldsymbol{x}_1^{(b)}\mid c_1,c_2\} \label{eqn_ass2}
\end{eqnarray*}
where $p(.)$ is general notation to denote a probability density or mass function.
\item[A.ii] $c_1$ and $c_2$ are independent: \{$p(c_1,c_2) =  p(c_1)p(c_2)$\}.
\item[A.iii] The unknown gene $c_2$ is not associated with ovarian cancer [$ p\{\boldsymbol{x}_1^{(b)}\mid c_1,c_2\} =  p\{\boldsymbol{x}_1^{(b)}\mid c_1\}$].
\item[A.iv] The penetrance $S_G(t\mid c_1)$, \textit{i.e.} the probability of developing disease by age $t$,  is known for breast and ovarian cancer in \textit{BRCA1/2} carriers. The prevalence $p(c_1)$ of \textit{BRCA1}, \textit{BRCA2} and the unknown gene are known in the population.
\item[A.v] A proportional-hazards model is assumed for the incidence associated with the unknown gene using two parameters $\boldsymbol{\theta} =(\beta, \gamma)$ through
\begin{eqnarray}
    S_G(t\mid c_1,c_2; \boldsymbol{\theta}) & = & S_0(t\mid c_1)^{\textrm{exp}(\gamma c_2)} \label{eqn_genhazmod}
\end{eqnarray}
        where $\gamma$ quantifies the log relative hazard attributable to the unknown gene, $\beta$ is its prevalence (the per-allele proportion is $1 - \sqrt{1-\beta}$), and $S_0(t\mid c_1)$ is a baseline survivor function.
       From assumption (A.iv) and equation (\ref{eqn_genhazmod}) the baseline survivor function is obtained by solving 
\begin{eqnarray}
    S_G(t\mid c_1; \boldsymbol{\theta}) & = & (1-\beta) S_0(t\mid c_1) + \beta S_0(t\mid c_1)^{\textrm{exp}(\gamma c_2)}, \label{eqn_baselinesurv}
\end{eqnarray}
        \textit{e.g.} via a Newton-Raphson iteration \cite{2004_TyrerCuzick}.
    \item[A.vi] Mendelian inheritance, \textit{i.e.} there is a 50\% chance that one of the two alleles from each parent is inherited by their offspring. 
\end{enumerate}
From these assumptions the breast cancer risk conditional on $\boldsymbol{x}_1$ is obtained as 
\begin{eqnarray}
S_G(t\mid \boldsymbol{x}_1;\boldsymbol{\theta}) & = & \sum_{c_1=0}^2\sum_{c_2=0}^{1}S_G(t\mid c_1, c_2;\boldsymbol{\theta})p(c_1,c_2\mid \boldsymbol{x}_1;\boldsymbol{\theta}) \label{eqn_mixmodel1}
\end{eqnarray}
where the weights $p(c_1,c_2\mid \boldsymbol{x}_1;\boldsymbol{\theta})$ in (\ref{eqn_mixmodel1}) are obtained from the following application of Bayes' rule. Denoting $\boldsymbol{d} = (c_1, c_2)$ then 
\begin{eqnarray}
p(\boldsymbol{d}\mid \boldsymbol{x}_1;\boldsymbol{\theta}) & = & \frac{p(\boldsymbol{x}_1\mid \boldsymbol{d};\boldsymbol{\theta})p(\boldsymbol{d};\boldsymbol{\theta})}{p(\boldsymbol{x}_1;\boldsymbol{\theta})}. \label{eqn_bayes1}
\end{eqnarray}
 $p(\boldsymbol{d};\boldsymbol{\theta})$ uses assumptions given in (A.ii, A.iv, A.v), and $p(\boldsymbol{x}_1\mid \boldsymbol{d};\boldsymbol{\theta})$ is obtained following A.vi (described in more detail in \cite{2004_TyrerCuzick}), first separately for breast $\boldsymbol{x}_1^{(a)}$ and ovarian cancer $\boldsymbol{x}_1^{(b)}$ family histories and then they are combined through (A.i). When gene testing has been done for \textit{BRCA1/2} the values for $c_1$ are taken from this, and the familial contribution only applied to the unknown gene $c_2$. Most of the computer code in the Tyrer-Cuzick algorithm is involved in calculating the likelihood $p(\boldsymbol{x}_1\mid \boldsymbol{d};\boldsymbol{\theta})$.

\subsubsection{Parameter estimates}

Some changes to the segregation model have been made since the original model \cite{2004_TyrerCuzick}, starting from version 7 onwards. \textit{BRCA1/2} prevalence and penetrance are now from the 1950+ birth cohort estimates reported in \cite{2008_Antoniouetal}. The penetrance estimate is now lower than used in original model \cite{2004_TyrerCuzick} (previously taken from \cite{1998_FordEastonetal}), which was attributed in \cite{2008_Antoniouetal} to bias in the original estimate \cite{1998_FordEastonetal} resulting from a focus on families with multiple cases of breast cancer. Prevalence of \textit{BRCA1} is taken to be 0.06\% and for \textit{BRCA2} is 0.10\% (previously 0.11\% and 0.12\% respectively). Smoothed first breast cancer rates from the Thames Registry 2005-2009 were used to calibrate overall risk, based on data reported by 5y age group and smoothed by  loess (with the smoother `span' parameter 0.2 chosen by eye). The unknown gene calibration has not been altered since the model was first introduced ($\beta=11.4$\% and $\textrm{exp}(\gamma)=13.04$, fitted to data  from \cite{2000_Andersonetal}, see \cite{2004_TyrerCuzick}). The impact of the unknown gene depends mostly on the product $\beta\textrm{exp}(\gamma)$ and it is difficult to accurately estimate these two parameters separately. The best fit gave a very large value for exp$(\gamma)$ but almost equally good fits would arise if $\gamma$ was substantially reduced and $\beta$ increased accordingly. Attributing risk to a single gene is a simplification, as risk is likely to reflect a combination of genes. However it is clear that the contribution of this unknown gene is greater than that for \textit{BRCA1} or \textit{BRCA2}, any other known gene or combination of single nucleotide polymorphisms and treating this either as a single unknown gene or a polygenic risk score has minimal impact on its predictive value. 

\subsection{Regression model}

\subsubsection{Model}
The relative hazard $r(\boldsymbol{x}_2)$ in model (\ref{eqn_phmod1}) is normalised by the mean risk in the population. This requires information on the relative hazard associated with other non-familial (personal, hormonal and lifestyle) risk factors \{denote this function $\phi(\boldsymbol{x}_2) > 0$\} and their population prevalence \{$f(\boldsymbol{x}_2)$ where $\int f(\boldsymbol{x}_2) \textrm{d}\boldsymbol{x}_2=1$\} in order to obtain the normalised relative hazard 
\begin{eqnarray}
    r(\boldsymbol{x}_2) & = & \frac{\phi(\boldsymbol{x}_2)}{\int \phi(\boldsymbol{x}_2)f(\boldsymbol{x}_2) \textrm{d}\boldsymbol{x}_2}.
\end{eqnarray}
In the Tyrer-Cuzick model this is approximated by treating each risk factor as independent (breast cancer risk factors in the model are largely independent of each other) and taking 
\begin{eqnarray}
    r(\boldsymbol{x}_2) & = & \prod_{j=1}^q \left\{ \frac{\phi(x_{2j})}{\int \phi(x_{2j})f(x_{2j}) \textrm{d}x_{2j}} \right\}.
\end{eqnarray}
When the value of a risk factor is unknown it can be left blank and the normalised relative hazard associated with that factor is taken to be 1 (woman assumed to have population risk). The mean risk constants used in the model are shown in Table \ref{ref-tbl1}.

\subsubsection{Parameters}
\label{sec-newpars}
The relative hazards and prevalence of risk factors in the original model were reported in \cite{2004_TyrerCuzick}, and included in Table \ref{ref-tbl1}. There have been changes and the current values are described below.

\begin{description}
    \item[Atypical hyperplasia or LCIS] The original model treated atypical hyperplasia or lobular carcinoma insitu (LCIS) as an independent risk factor. Evidence in \cite{2007_Degnimetal} and \cite{2010_Bougheyetal} indicated that other risk factors should not be included when a women has been diagnosed with atypical hyperplasia as it is more of an intermediate endpoint. As a result we modified the model to take the maximum risk of atypical hyperplasia (only), LCIS (only) and family history risk combined with other risk factors (based on the 10y risk). For example, a young \textit{BRCA1/2} carrier with atypical hyperplasia would not have her risk assessment modified by atypical hyperplasia. This is likely to be conservative, but currently there is inadequate data to model the joint effects of atypia and other risk factors accurately.
    \item[Unknown benign disease] In general the risk of subsequent breast cancer associated with a benign lesion depends on the histology, with no increased risk associated with non-proliferative lesions, about a two-fold risk associated with hyperplasia of the usual type and about a 4-fold risk associated with atypical hyperplasia \cite{2018-dupont}. Based on prevalence data reported by \cite{2018-dupont}, the relative hazard from benign disease when a women has had a biopsy but pathology is unknown is taken to be 1.3; this option was not available originally. Non-biopsied lesions are not considered in the model. 
    \item[Menopause hormone therapy] The model assumes a maximum relative hazard of 2.0 for combined therapy with an oestrogen and progestin and 1.4 for oestrogen-only hormone therapy. The relative hazard begins at unity in year 1, is half the maximum excess relative hazard in year 2 and then the maximum relative hazard therafter until stopping therapy. Risk is then ramped down following cessation to be 2/3 of the maximum excess in the first year after stopping, then one third in the next year, and no increase thereafter. The relative risk is adjusted if BMI (body mass index) is known, being decreased by 10\% of the average excess risk if obese (BMI $\geq$ 30 kg/m$^2$), and increased by 10\% if BMI $<$25 kg/m$^2$ (not overweight or obese).
        These assumptions are based on results from the Women's Health Initiative trial \cite{2011_ChlebowskiAnderson}) and the Million Women Study \cite{2006_Reevesetal}.

\end{description}

\subsection{New additions to the model}

\subsubsection{Mammographic density}

\begin{figure}[h]
          \begin{center}
                      \includegraphics[scale=0.7]{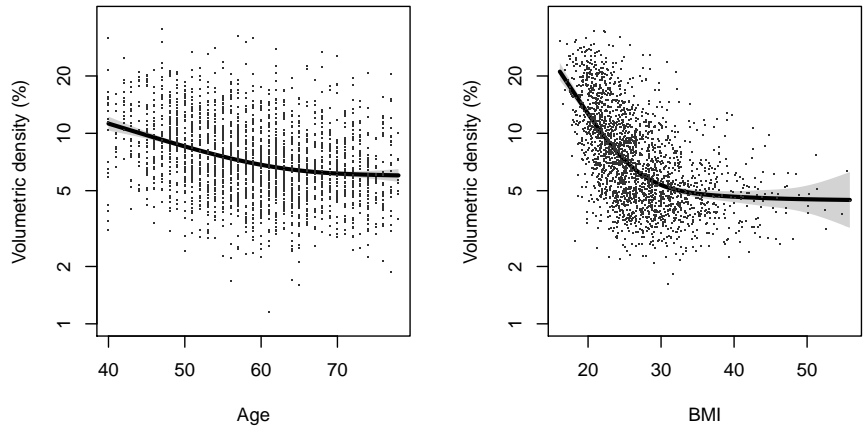}
                          \caption{Volumetric percentage density vs age and body mass index (BMI), with joint non-parametric smooths (line ---). Taken from \cite{2019-brentnallharveyetal}.}
                                    \label{fig_volpara}
                                      \end{center}
                                            \end{figure}

Breast density is correlated with some of the other risk factors in the model, most strongly age and body mass index. It is also measured in different ways that require different calibration. We allow three different ways to input density: a visual assessment in which (ideally) two readers estimate the percentage of the breast covered by dense (opaque) tissue and the average is used, the BI-RADS system which corresponds to four categories, and an automated volumetric system (Volpara) that estimates the percentage volume of dense tissue based on the radiographic intensity of each pixel. In order to incorporate breast density into the model without changing the effect sizes of the other factors, we first developed a measure of breast density which is independent from age and BMI at mammogram, by taking the difference between observed and expected density given age and BMI. This was developed for the different measures in case-control studies, where expected density was modelled in controls by fitting a generalized additive model \cite{2017-wpod} of different breast density measures against splines for age and BMI (see Figure \ref{fig_volpara} for a fit to volumetric percent density). The risk associated with density was then calibrated by estimating the effect of the density residual, adjusted for other factors in the Tyrer-Cuzick model using case-control and cohort studies \cite{2014_Warwicketal,2015_BrentntallEvansetal,2018-brentnalletal}.

\subsubsection{Polygenic risk scores}

Numerous studies have demonstrated the utility of polygenic risk scores. These  use common gene variants which individually carry small risks but as a group provide useful information. Their utility has been observed in average risk women \cite{2010_Wacholderetal, 2010_Mealiffeetal} and in high-risk women \cite{2016-cuzickbretnall-jco,citeulike:14246576}, and they combine effectively independently with classical risk factors and breast density \cite{2015_Vachon2etal,2018-vanveenetal}. Currently, the breast cancer association consortium has validated more than 170 risk-modifying SNPs at genome-wide significance \cite{2018-snp313}, and this is likely to increase still further.

To provide an overall relative risk estimate from a collection of independent SNPs polygenic risk scores may be obtained by using published data on their per-allele risks, and allele frequencies. First the odds ratio for each of the three SNP genotypes is determined (no risk alleles, 1 risk allele, and 2 risk alleles) and these are normalised to obtain an average of unity using reported risk allele frequencies. In our model we assume the odds of two copies of the abnormal gene is the square of that for one copy, and their distribution follows the Hardy-Weinberg law. Finally to obtain an overall polygenic risk score the odds ratios for each of the genotypes are multiplied together assuming independence \cite{2010_Mealiffeetal}.

Data show very small correlations between polygenic risk scores and classical risk factors. We have observed that an accurate fit is obtained by treating the score as independent of the other factors in the Tyrer-Cuzick model \cite{2016-cuzickbretnall-jco,citeulike:14246576,2018-vanveenetal}. However, theoretically, a polygenic risk score will explain some of the genetic aggregation modelled through the segregation model, and so this should be down weighted in the risk assessment. Work to refine this component is ongoing.

\section{Comparison with other risk models}

It is important to understand differences between risk models, both for use of the models and to refine the scientific process of improving risk assessment. Here we note some differences between the Tyrer-Cuzick and other models. 

\subsection{Segregation models}

Some segregation models, including BRCAPRO \cite{1998-parmigiani}, calculate breast cancer risk only based on estimating the likelihood of carrying major genes based on a personal and family history of breast, ovarian or other cancers related to these genes. The main aim of such models is to estimate risk of being a \textit{BRCA1/2} mutation carrier. They are mostly used for women with a strong family history in order to determine eligibility for \textit{BRCA1/2} testing. They incorporate both pedigree information including age at onset and results of genetic testing for \textit{BRCA1} and \textit{BRCA2} in the women and their relatives. Such models are less suitable for risk assessment of breast cancer in the wider population, partly because they do not allow for other unknown genetic factors in the model. 

The BOADICEA model accounts for other rare but highly penetrant genes  (\textit{PALB2}, \textit{CHEK2}, \textit{ATM}), and uses a polygenic model to allow for (unobserved) risk due to other genetic factors than those measured \cite{2016-leentonioetal}. Define $c^*_1=0$ if not positive for one of the genes, $c^*_1=1$ if test positive for \textit{BRCA1}, $c^*_1=2$ if test positive for \textit{BRCA2} (but not \textit{BRCA1}), $c^*_1=3$ if test positive for \textit{PALB2} (but not \textit{BRCA1/2}) and so on for next \textit{CHEK2} and last \textit{ATM}. Thus very rare individuals who test positive for more gene are assigned a group $c^*_1 = 0,\ldots,5$ based on a hierarchy. The model has the following form for the breast cancer specific hazard:
\begin{eqnarray*}
    h_1\{t \mid c_1, u(t)\} & = & h_{1B}(t) \textrm{exp}\left\{ \sum_{j=1}^5 \beta_j(t) I(c^*_1=j)  + u(t)\right\},\label{eqn_mod1}
\end{eqnarray*}
where $t$ is age, $h_{1B}(t)$ is a baseline hazard function, $I(.)$ the indicator function, $\beta_j(t)$ are time-dependent relative hazards for each genotype, and $u(t)$ is a time-dependent random effect from a normal distribution with mean zero and standard deviation $\sigma_t$. When the woman has not been personally tested, the probability of being a carrier is estimated based on the pedigree and testing results (if any) in family members. The random (polygenic) effects arise from a combined effect of several unmeasured and largely unknown genes, each of which contributes a small effect. Large scale genome-wide association studies have identified that the polygenic model for genetic susceptibility is partly correct, but a large proportion of familial aggregation of risk is still not explained by known major genes or polygenic risk \cite{2018-snp313}. 

For most pedigrees arising in the general population one might expect only small differences for risk estimates from polygenic \textit{vs} single unknown gene models.  Qualitatively, the unknown single dominant gene model is a good approximation to the polygenic model: the genotype is never observed and so gives rise to a range of effects through equations (\ref{eqn_mixmodel1}) and (\ref{eqn_bayes1}). Indeed, one study that compared the fit between the unknown dominant gene and polygenic models did not show an appreciable difference between them when considering additional genetic factors other than \textit{BRCA1/2} \cite{2002-boadicea}. 

Another difference between BOADICEA and Tyrer-Cuzick is that the hazard for the unknown genetic effect $\gamma$ is assumed to not vary with age in the Tyrer-Cuzick model, whereas the polygenic variance varies with age in the BOADICEA model. Thus the BOADICEA model leads to a larger decline in the relative risk as the age of the affected relative increases and appears to fit better with the literature. We will update this aspect of the model in the next version. 

\subsection{Regression function models}

Several models have been developed based on a regression function only. These include the Gail (or BCRAT) \cite{1989_Gailetal}, the Breast Cancer Surveillance Consortium (BCSC) \cite{2008_Ticeetal,2015_ticeetal} and Rosner-Colditz models \cite{2017-rice,Zhang2018}. Neither the Gail nor BCSC model include all the risk factors included in the Tyrer-Cuzick model, but they do account for ethnic differences. The Rosner-Colditz model includes the terms for the same broad risk factors as in the Tyrer-Cuzick model, and some additional factors including alcohol consumption, adolescent body somatotype (shape / size) and hormone levels. Assumptions on the risks and prevalence of risk factors differ between the models, as do the use of interaction terms. 

The Tyrer-Cuzick model includes an interaction between menopausal status and BMI, assuming it is only a postmenopausal risk factor. Some data suggest that BMI may actually be a protective factor for premenopausal breast cancer but this is not included in the model \cite{Schoemaker2018}. Interactions are also included for BMI and HRT (as discussed above), and atypical hyperplasia (or LCIS) and all other risk factors in that it is treated as an intermediate endpoint, and the maximum risk for it alone or all other factors combined is used, following analysis from \cite{2010_Bougheyetal}. No other interaction terms are currently included.

The Gail model includes an interaction between (i) ethnicity and all other risk factors, and for certain ethnic groups an interaction between (ii) age and number of biopsies, and (iii) age at first child and number of affected first-degree relatives (negative interaction). For example, both latter interactions are included for white women, but they are not used for Asian women. 

The BCSC model includes interactions between a woman's age with (i) ethnicity, (ii) benign disease, (iii) family history (negative interaction) and (iv) breast density (negative interaction), by allowing for different regression coefficients in 10y intervals starting at age 40y \cite{2015_ticeetal}.

The Rosner-Colditz model allows for interactions between: (i) BMI and duration of premenopausal period (time from menarche to start of menopause) (ii) BMI and current duration of menopausal interval, (ii) alcohol after menopause and hormone replacement therapy, (iii) benign disease with age at menarche, (iv) benign disease with duration of premenopause, (v) benign disease with duration of menopause, (vi) height with duration of premenopause (vii) height with duration of postmenopause. 

All the models (except Tyrer-Cuzick) include interactions relating to benign disease, but their forms are quite different. In the Gail model, for white women the relative risk of previous biopsies is less in women older than 50y than for younger women. In the BCSC cohort used to fit their model \cite{2015_ticeetal} this pattern was not observed for proliferative disease, and for non-proliferative disease the direction was even reversed (a larger effect for women older than 50y). In the Rosner-Colditz model there is a positive interaction with age at menarche and a negative interaction for the other terms (albeit none are univariately significant in the most recent model at a 5\% level \cite{2017-rice}), suggesting an older age at menarche is not protective for women with benign disease.  These three models have been fitted in populations with individual-level data. Relative hazards for white women in the Gail model were estimated using a case-control study with 2842 cases and 3146 controls, including 179 cases with a prior biopsy aged $<$50y and 487 cases with a prior biopsy aged 50y+ \cite{1989_Gailetal}. The BCSC model was fitted to a cohort with more than 1 million women with up to 10y follow up, and included 6204 women with proliferative disease without atypia and 177 subsequent cancers \cite{2015_ticeetal}. A recent Rosner-Colditz model update included approximately 100,000 women of whom 5,246 had developed breast cancer. Benign breast disease (non-proliferative and proliferative) was reported in approximately 18,000 women at entry \cite{2017-rice}.  

Potential interactions between benign disease and other risk factors have also been investigated in other studies, but also with inconclusive findings. For example, in a cohort from Nashville a negative interaction between non-proliferative disease and age was observed, such that women diagnosed with non-proliferative disease at an older age had lower breast cancer risks than women in the cohort without proliferative disease \cite{2018-dupont}; this has not been evaluated in any of the four models considered here. 

Overall, due to the current lack of strong evidence to support an interaction of benign disease with age, no such interaction is included in the Tyrer-Cuzick model, and it assumes non-proliferative benign disease does not confer any increased risk.

\section{Absolute cumulative risk assessment and calibration}
\label{sec:oeoverall}
\subsection{Framework}

It is important to be aware of the difference between using a model to project risks and calibrating a model from a cohort where follow-up data already exists. In the former case routine data only provides risk assuming no inter-current mortality, and a specific adjustment needs to be made for competing mortality (\textit{c.f.} equation \ref{eqn:absrisk}). In the latter case, inter-current deaths are known and can be treated as part of the censoring process. This has led to different methods for computing absolute cumulative risk, which are essential for interpreting the model predictions, and subsequently to assess how well the model is calibrated to the population under study. This has not been widely appreciated, and can lead to different estimated values of expected absolute risk from the same disease model. In this section we review some different methods for computing expected cumulative risk under the following setup. 

Assume there is a sample of $i=1,\ldots,n$ independent individuals with data on risk factors. The risk model evaluates each risk from current age $t_0$ to age $t$, defined as  $P_{x_i}(t_{0i}, t)$ or $P_i(t)$ using shorthand notation following equation (\ref{eqn:absrisk}). The time to breast cancer or death is subjected to right-censoring, so that one observes $\breve{T}= \textrm{min}(T_{1}, T_{2}, T_C)$, where $T_1$ is time to disease, $T_2$ is the time to death from other causes and $T_C$ is a right-censoring time. 

An issue to contend with for computing the expected probability of breast cancer is that the `at-risk' interval should be from $t_0$ to $T_C$, whereas $\breve{T}$ is often used. $T_C$ is not observed when $\textrm{min}(T_1, T_2) <T_C$ and this presents difficulties as its (conditional) distribution can be hard to determine. Calculating the expected risk using probability of breast cancer over $(t_0, \breve{T})$ is liable to underestimate expected risk. A solution is to use lifetable methods which are based on the sub-hazard of breast cancer and not the probability of breast cancer, but only need to be computed over $(t_0, \breve{T})$. We next consider these different methods in more detail.

\subsection{Expected risk based on the cumulative disease-specific hazard function }
\label{sec-expect-cumH}

The expected number of breast cancer cases in the cohort may be obtained by first integrating breast cancer specific hazards over the observed follow-up period, to obtain each individual's cumulative hazard for breast cancer during the period she is at risk. 
Formally, let $Y_i(t) = I(\breve{T}_i>t)$ denote the at-risk process for individual $i=1,\ldots,n$, and $h_1(t \mid \boldsymbol{x}_i) = h_{1i}(t)$ the disease-specific hazard function sub-model. Then the expected number of individuals with disease ($J=1$) is 
\begin{eqnarray}
    {E}^{(H)} & = & \sum_{i=1}^n \int_{t_{0i}}^\infty h_{1i}(u) Y_i(u)\textrm{d}u \label{eqn:neta} \nonumber \\
    & = & \sum_{i=1}^n    H_{1i}(\breve{T}_i) \label{eqn:net}
\end{eqnarray}
where \begin{eqnarray*}
    H_{1i}(t) & = & \int_{t_{0i}}^{t} h_{1i}(u) \textrm{d}u.
\end{eqnarray*}
Equation (\ref{eqn:net}) may be adapted to compute the expected number of deaths from other causes by replacing $h_1$ with $h_2$, etc. Indeed, by letting $h_2$ reflect all-cause mortality (including breast cancers) this is equivalent to the time honoured lifetable method to compute the expected number of deaths \cite{1987-keiding}. Thus while it might seem surprising to base the expected number of women diagnosed with breast cancer on the disease-specific cumulative hazard rather than the smaller estimate of the probability of breast cancer, this method is firmly established in lifetable and other settings (\textit{e.g.} the log-rank test). 

\subsection{Expected risk based on cumulative incidence}
\label{sec:cumincexp}
The previous section is the preferred method (see below) but another approach is to calculate expected risk using the model probability of disease allowing for death from other causes (cumulative incidence). This is most straightforward when the model is used in a cohort with no censoring from starting age $t_0$ up to age $t$. In this case the expected number is a summation of the conditional cumulative incidence in (\ref{eqn:absrisk}):
\begin{eqnarray}
    E^{(P)}(t) & = & \sum_{i=1}^n P_{i}(t), \label{eqn:crudefix}
\end{eqnarray}
where $P_i(t) \equiv P_x(t_0, t)$ for the $i$-th individual. It is extremely uncommon to have such a dataset, and the calculation is more complicated under the most common scenario of right censoring because we need an estimate of $\textrm{P}\{J=1, \textrm{min}(T_{1}, T_2, T_C) \leq t \mid \boldsymbol{x}_i\}$, whereas $P_i(t)$ estimates $\textrm{P}\{J=1, \textrm{min}(T_{1}, T_2) \leq t \mid \boldsymbol{x}_i\}$ (see equation \ref{eqn:absrisk}). We consider three types of censoring processes for prospective risk calculation.  

Firstly, if censoring is fixed to be to a common follow-up time for each patient $t_i=t_{0i}+A$, such as $A=5$ years, then 
\begin{eqnarray}
    E^{(P)}(A) & = & \sum_{i=1}^n P_{i}(t_{i0}+A). \label{eqn:crudefix0}
\end{eqnarray}
Secondly, when the censoring time $T_{C}$ is variable between individuals but it is deterministic, such as being due to variable calendar time enrolment and a common follow-up for all $i$ then 
\begin{eqnarray}
    E^{(P)} & = & \sum_{i=1}^nP_{i}(t_{i0}+T_{Ci}). \label{eqn:crude}
\end{eqnarray}
A third case is when the potential censoring time is unknown but is a stochastic process with known distribution. Then one possibility is to base expected risk on 
\begin{eqnarray}
    P_{C}(t) & = & \nonumber 
    \\ & & \int_{t_0}^{t_0+t} h_1(u \mid \boldsymbol{x}) \textrm{exp}\left[ - \int_{t_0}^{t_0+u} \{h_1(v \mid \boldsymbol{x}) + h_2(v \mid \boldsymbol{x})\} \textrm{d}v \right]{S}_C(u \mid \boldsymbol{x})\textrm{d}u \label{eqn:absrisk22b} 
\end{eqnarray}
where ${S}_C$ is the survivor function for the censoring distribution. Then
\begin{eqnarray}
    E^{(P)} & = & \sum_{i=1}^n P_{Ci}(t_{i0}+\breve{T}_i).  \label{eqn:crude2}
\end{eqnarray}
If $S_C$ is unknown then one will need to estimate it, for example by assuming censoring is independent of risk factors $\boldsymbol{x}$ and using the Kaplan-Meier estimator \cite{2002_Lawless}. 

\subsection{Calibration}
\label{sec:basiccalib}
The above methods based on cumulative disease-specific hazards or cumulative incidence may be used to assess calibration by comparison of the observed number of women with disease ($O$) with expected ($E$), both overall and in sub-groups \cite{2013-hoslem}. If it is assumed that disease is rare (this assumption is relaxed in Section \ref{sec-extensions}), then a test may be constructed assuming that $O$ is generated from a Poisson distribution, whose rate would be $E$ if the model was perfectly calibrated. Exact Poisson confidence intervals on $O$ may be used to compute a confidence interval for $O/E$ (treating $E$ as fixed) and determine if it covers unity \cite{2005-gailpfieffer}.

\subsection{Some biased estimates of absolute risk and their effect on calibration}
\label{ref:errorsavoid}
Some commonly-used methods are biased, but this can be overcome using hazard-based methods where the appropriate `at-risk' interval is used. We next consider three methods that yield biased estimates of $E$, the expected number of cancers, which in turn leads to biased estimates of the observed to expected ratio $O/E$.
\begin{enumerate} 
    \item Under right censoring some studies \cite{1999_Costantinoetal,2006-decarlietal,2012_banegas,2013-Rosneretal} have used cumulative incidence as the measure of expected risk, but estimated it via 
\begin{eqnarray}
    \sum_{i=1}^n P_i(\breve{T}_i) \label{eqn:sumtoT}.
\end{eqnarray}
        This measure of expected risk is biased towards zero because it only computes risk until the event time $\breve{T}$. For example, suppose we wish to compute the expected number of deaths in a sample of babies over the next 200y. If the mortality model is well calibrated but risk of death until the age at which each person actually dies is summated (as equation \ref{eqn:sumtoT}), then fewer deaths will be expected than the number of babies in the sample. The correct analysis based on cumulative incidence would summate risk of death by age 200y for each individual (\textit{c.f.} equation \ref{eqn:crudefix}). In mitigation, the bias will be small, and will have a minimal impact on the conclusions drawn when the event is rare. 

    \item In some studies it has been reported \cite{2003_Amiretal} that the expected number with disease is computed as 
\begin{eqnarray}
    \sum_{i=1}^n [1 - \textrm{exp}\{-H_{1i}(\breve{T}_i)\}], \label{eqn-wrongagain}
\end{eqnarray}
        \textit{i.e.} summating 1 minus the net survival (the estimand for Kaplan-Meier estimation) rather than (\ref{eqn:net}). In general this is biased towards zero for the expected number of breast cancers because the term in the summation is less than or equal to the cumulative hazard $H_{1i}(\breve{T}_i)$. However, this is a small bias if the cumulative hazard $H$ is small since $1 - \textrm{exp}(-H) = H + O(H^2)$. Also note that $1 - \textrm{exp}\{-H_{1i}(\breve{T}_i)\} \geq P_i(\breve{T}_i)$, so is less biased than summating $P_i(\breve{T}_i)$ (\textit{i.e.} equation \ref{eqn:sumtoT}). 

    \item Finally, calibration of breast cancer risk has sometimes been assessed using a fixed follow-up time for all by only including those who could be followed up at least that long \cite{2008_Ticeetal,2001-rockhilletal}. This in inefficient because some data would be excluded, and it should be used with caution when there is censoring. For example, suppose we seek to assess calibration of 5y risk when there is right censoring in the data, and include only those without breast cancer who were not censored by 5y and all breast cancer cases diagnosed up to $5$y. If the model is correct (and censoring is non-informative) then $O/E$ based on (\ref{eqn:crudefix}) will be biased. $E$ will be too small relative to $O$ because some of the non-cases will have been excluded from expected risk due to censoring, but they should have contributed to $E$ as they would have been counted as cases if disease had occured in them.
\end{enumerate}

\subsection{An unbiased estimate of expected risk}
\label{sec:recoA}

To avoid bias we recommend using $E^{(H)}$ to obtain the expected number with disease. This is based on the cumulative hazard rather than probability of disease (cumulative incidence) and has several advantages compared to $E^{(P)}$ when censoring has been handled appropriately as described in Section \ref{sec:cumincexp}. 

Firstly, comparing $E^{(H)}$ with $O$ is a test of whether the disease-specific hazard model is correct. The same method may also be used to test the competing-mortality model. Comparing $E^{(P)}$ with $O$ is a test of whether the model for cumulative incidence based on combining the disease risk and competing mortality is correct, \textit{i.e.} whether the marginal risk based on (\ref{eqn:absrisk}) is well calibrated. This might require accurate knowledge and data on the competing risk process, which is often not available.  It is uncommon for the competing mortality model to be as well developed as the disease risk model, which is of primary interest to the epidemiologist. If the competing mortality model is incorrect, but the disease model is correct, then the expected risk will be incorrect due to an inadequate model for competing mortality. Conversely, even if the cumulative incidence function risk is well calibrated, it is possible that the competing mortality and disease model are both wrong in different ways, and just happen to cancel each other out. Comparing the hazard sub-models separately would identify such issues.

Secondly, when there is stochastic censoring equation (\ref{eqn:net}) is easier to apply than (\ref{eqn:crude2}) because it does not require an assumption or estimate for the conditional censoring distribution.

Thirdly, it can be a more powerful test. For example, in an extreme example of 200y follow-up on mortality in babies, then one would not be able to show that a model that assigns a constant and excessively high mortality rate was incorrect; whereas analysis based on summating the cumulative hazard of that model to the time of each death would yield an expected number of deaths in excess of observed.

\subsection{Extensions}
\label{sec-extensions}
We end this section by describing methods to assess calibration not only for the total population, but also in subgroups. For example, when the data are split into ten groups by decile of the predicted risk at baseline then a test of calibration (more generally model fit) across the groups would be akin to the Hosmer-Lemeshow test for binary outcomes \cite{1996-harreletal,2013-hoslem}. To extend this strategy to the cumulative incidence framework, we construct $k=1,\ldots,K$ distinct groups, and assess whether the observed risk matches expected in each of the groups. Then, the test statistic
\begin{eqnarray*}
    \chi^2_{K-1} & = & \sum_{k=1}^K \frac{(O_k - E_k)^2}{E_k}, \label{eqn:riskcalib}
\end{eqnarray*}
where the expected $E_k$ are obtained as above, will be approximately $\chi^2$ with $K-1$ degrees of freedom under the null hypothesis of correct prediction and calibration. 

This hypothesis testing approach has limitations. Firstly, power to reject inaccurate calibration is limited by the large number of degrees of freedom. Secondly, with a large enough sample size we would expect to reject model calibration because all models are wrong, and the approach does not directly indicate where the inaccuracies lie. 

An alternative approach is to view testing calibration as special case of Poisson regression. Let $O_{i}$ denote whether individual $i=1,\ldots,n$ in the sample has disease, and $E_{i}$ be the expected risk based on the model. Then consider a Poisson regression for
\begin{eqnarray*}
    \textrm{E}(O_{i} \mid E_{i}) & = & \textrm{exp}\{\gamma_0 + \gamma_1 \textrm{log}(E_{i})\},  \label{eqn:possreg2}\\
     & = & \theta E_i^{\gamma_1}
\end{eqnarray*}
where $\theta = \textrm{exp}(\gamma_0)$ and $\gamma_1$ are unknown parameters. Setting $\gamma_1=1$ then the maximum likelihood estimate $\hat{\theta} = \sum_i O_i / \sum_i E$ provides an overall `calibration-in-the-large' parameter for the entire cohort \cite{2016-crowsonetal}. When $\gamma_1$ is also estimated then it can be used to test for calibration across the range of expected risk (on 1 df), as well as yielding an estimate of how closely the regression line matches observed ($\gamma_1=1$ being ideal; 0 being not at all).

Finally, we consider when the disease is relatively common, perhaps because focus is on a high risk group with long follow up. In this case the Poisson distribution no longer approximates a binomial distribution. However, Poisson regression may be valid if time is broken into shorter segments where the rare disease assumption holds, for example by year, with each segment treated as a `new' observation in the Poisson regression analysis. This approach is also useful when assessing time-dependent calibration, and which we proceed to consider next.

\section{Time-dependent calibration} 
\label{sec:tdcalib}
The focus of many studies has been on whether the cumulative number with disease over a single follow-up period matches the expected number. While important, it does not assess differences through time, such as non-proportionality of the observed to expected risk. In this section we consider techniques to assess calibration across follow-up time, and develop new graphical methods.

\subsection{Methods}

One approach is simply to look at the observed and expected number of events $N_j(t)$ for each cause $j$ as a counting process. The observed number is
\begin{eqnarray}
	\hat{N}_j(t) & = & \sum_{\breve{T}_i \leq t} I(J_i=j).\label{eqn:nelsona}
\end{eqnarray}
The expected number, based on the cumulative hazard approach (as Section \ref{sec-expect-cumH}), is
\begin{eqnarray}
    {N} _j(t) & = & \sum_{i=1}^n \int_0^t Y_i(u) h_{ji}(u) \textrm{d}u \nonumber \\
    & = & \sum_{\breve{T}_i \leq t} H_{ji}(\breve{T}_i) + \sum_{\breve{T}_i > t} H_{ji}(t), \label{eqn:nelsona2}
\end{eqnarray}
so that we have extended (\ref{eqn:net}) to be a function of $t$. One may compare observed $\hat{N}_j(t)$ \textit{vs} expected from (\ref{eqn:nelsona2}) through follow-up time $t$, such as is sometimes done to check proportional hazards  \cite{arjas1988,Gronnesby1996}. Cumulative pointwise (such as from entry to time $t$), or interval (such as number in a year) confidence intervals may be constructed using exact Poisson confidence intervals, or used to undertake hypothesis tests of calibration up to or near time $t$.

An issue with the comparisons based on $N(t)$ is that they depend on the sample size and censoring distribution, so cannot be readily used to compare between studies. To circumvent this one might consider the comparing the mean cumulative hazard among those at risk, \textit{i.e.} for $t < \textrm{max}_i(\breve{T}_i)$ comparing 
\begin{eqnarray}
	\hat{H}_j(t) & = & \sum_{\breve{T}_i \leq t} \frac{I(J_i=j)}{Y_+(\breve{T}_i)} \label{eqn:nelson}
\end{eqnarray}
with its expected value
\begin{eqnarray}
    {H}_j(t) & = & \sum_{i=1}^n \int_0^t \frac{Y_i(u) h_{ji}(u)}{Y_+(u)} \textrm{d}u \label{eqn:nelson2}
\end{eqnarray}
where the total number at risk at time $t$ is $Y_+(t) = \sum_{i=1}^n Y_i(t)$. 
Note that equation (\ref{eqn:nelson}) is the Nelson-Aalen estimator of the cumulative hazard \cite{2002_Lawless}. Equation (\ref{eqn:nelson2}) is the expected cumulative hazard based on the breast cancer risk model and observed at-risk process.
Time-dependent assessment of the observed to expected risk can be based on $\hat{H}_j(t)$ \textit{vs} ${H}_j(t)$ across follow-up time $t$. Here we treat $H_j(t)$ as fixed and use the usual confidence intervals for the Nelson-Aalan estimator $\hat{H}_j(t)$ at different times, from which the hypothesis of calibration at a particular time $t$ can also be assessed. 

Other functionals of $\hat{H}_j(t)$ and $H_j(t)$ might also be used for assessment of calibration across time. Arguably the most common functional used in applications is the `net' risk
\begin{eqnarray*}
    S(t) & = & 1-\textrm{exp}\{-H(t)\}, \label{eqn-netrisk}
\end{eqnarray*}
which is often estimated as 1 minus the Kaplan-Meier estimator \cite{2002_Lawless}. There are two options for calculating the expected risk. The first is a mean from the entire cohort at baseline
\begin{eqnarray}
    S^{(A)}_j(t) & = & n^{-1} \sum_{i=1}^n [1-\textrm{exp}\{-H_{ji}(t)\}]. \label{eqn-pt1}
\end{eqnarray}
The second is to use the mean hazard in those still at risk at time $t$ through
\begin{eqnarray}
    S^{(B)}_j(t) & = & 1 - \textrm{exp}\{-H_j(t)\} \label{eqn-pt2}
\end{eqnarray}
and (\ref{eqn:nelson2}). 

A disadvantage of the cumulative hazard (\ref{eqn:nelson2}) and net risk (\ref{eqn-pt1},\ref{eqn-pt2}) is that interpretation of the component terms is not straightforward in relation to the observable risk in equation (\ref{eqn-sub}). However, if we weight (\ref{eqn:nelson}) by the proportion at risk using the Kaplan-Meier estimator $\hat{S}(t)$ for all-cause survival (\textit{i.e.} the survival endpoint is disease or death), then we recover an estimator of the cumulative incidence function through 
\begin{eqnarray}
	\hat{F}_j(t) & = & \sum_{\breve{T}_i \leq t}  \frac{I(J_i=j)}{Y_+(\breve{T}_i)} \hat{S}(\breve{T}_i),  \label{eqn:cif}
\end{eqnarray}
and inference may be based on estimates of its variance \cite{2002_Lawless}. 
Expected risk may be obtained from  
\begin{eqnarray}
    F_j^{(A)}(t) & = & n^{-1}\sum_{i=1}^n P_i(t), \label{eqn:cif2a}
\end{eqnarray}
or the mean expected risk at each $t$ over the $n$ individuals.

\subsection{Summary}

Our recommendation is to focus on comparisons only involving the disease-specific hazard to assess $O$ vs $E$ through follow-up time $t$. For exploratory plots we prefer use of the cumulative hazard comparison of (\ref{eqn:nelson2}) with (\ref{eqn:nelson}). There are several reasons for these recommendations.

Firstly, as for overall calibration, the method may be used to assess the disease risk and competing mortality sub-models separately. The cumulative hazard analysis extends quite naturally a total number of events comparison, but avoids dependence on the censoring distribution.

Secondly, focusing on the hazard leads to using the observed at-risk process to calculate expected risk, which appears to be a more robust method to assess model calibration than only using a risk assessment estimate on everyone at baseline. For instance, if those censored or lost to follow-up are more likely to be at higher predicted risk at entry, then estimates of their expected risk will differ, and so comparison of expected risk between methods based on using the at-risk-process or everyone at baseline is a way to assess whether the censoring process was associated with predicted risk. In general terms this issue has parallels with relative survival, where one might consider the Ederer-I approach (mean risk) or the Ederer-II approach (use the at-risk process) \cite{2017-sasienibrenntall}.

Thirdly, a statistical advantage of the cumulative hazard compared to the cumulative incidence function is that it may have a smaller variance because it does not require an assumption or estimation of all-cause survival (equation \ref{eqn:cif}).

Finally, in the exploratory plots when the aim is to assess calibration of a risk model, then differences between the observed to expected cumulative hazard seem easier to interpret than the other methods. In contrast, the cumulative incidence function weighs the cumulative hazard by all-cause survival. The component terms have a natural interpretation, but the observed to expected comparison is perhaps less easy to interpret for assessing calibration than cumulative cause-specific hazards, because changes could be due to either lack of calibration in the disease or competing mortality models. Kaplan-Meier estimation transforms the cumulative hazard  through $1-\textrm{exp}(-H)$, so that differences between large cumulative hazards are downweighted. This makes changes in the calibration plot with time less easy to relate to calibration at the hazard level. For example, it is easier to assess whether the hazard function is constant through time by plotting the Nelson-Aalan estimate of the cumulative hazard, than by plotting 1 minus the Kaplan-Meier estimator.

\section{Example}

\subsection{Data}

We have previously reported an evaluation using the Tyrer-Cuzick model in a cohort from Washington state USA \cite{2018-brentnalletal}. Between 1996 and 2013, 132 139 women aged 40-73y completed a risk questionnaire and had a measure of breast density taken at entry. They were followed up beginning 6-months after the entry mammogram to the earliest of diagnosis of invasive breast cancer or censoring. Women were censored due to disenrollment (n=62 331, 48.2\%), end of follow-up (n=48 317, 37.3\%), age 75y (n=15 827, 12.2\%), death from other causes (n=2328, 1.8\%) or ductal carcinoma \textit{in situ} (n=637, 0.5\%). Only aggregate data on competing risk causes were made available for analysis, so we were unable to apply methods that require individual-level data about the cause of censoring. 

\subsection{Calibration}
\label{sec:data:calib1}
A total of 2699 breast cancers were observed ($O$), and based on our preferred method (equation \ref{eqn:net}) 2757 were expected ($E$), yielding $O/E$ 0.98 (95\%CI 0.94-1.02). We next consider application of incorrect methods discussed in section \ref{ref:errorsavoid}.

A first biased assessment is cumulative incidence over a 5y fixed-time horizon by using all the cases that were diagnosed in the 5y period and all non-cases at risk for 5y or more. It we calculate the expected 5y risk in those who are at risk at 5y then $E=877$ against an observed $O=1157$ (O/E = 1.32). A second biased method is to summate expected cumulative incidence to the last follow-up for each individual (equation \ref{eqn:sumtoT}), yielding $E=2605$ and $O/E = 1.04$ (95\%CI 1.00 to 1.08). A third biased method is to follow equation (\ref{eqn-wrongagain}) and summate the absolute net risks, which gives expected number $E=2679$ (O/E = 1.00). 

It also instructive to repeat the above methods in a high-risk group ($>8\%$ ten-year net risk) where greater differences between the methods to calculate $E$ will occur. The Tyrer-Cuzick model with mammographic density identified 4645 women to be in this group, and 273 breast cancers were subsequently diagnosed. Based on the preferred method (\ref{eqn:net}) we find $E=349$, so that $O/E=0.78$ (95\%CI 0.69-0.88), indicating that the model over predicted the high risk group. If instead (\ref{eqn-wrongagain}) was used then $E=324$ and $O/E=0.84$ (95\%CI 0.75-0.95). Based on (\ref{eqn:sumtoT}) we find $E=310$ so that $O/E=0.88$ (95\%CI 0.78-1.00), where the 95\%CI covers unity. Thus the correct analysis showed a lack of calibration that would not be seen so clearly with the biased methods.

This analysis shows the method to calculate expected risk can cause practically important differences in interpretation of risk model calibration, particularly for high-risk groups. 

\subsection{Time-dependent calibration plots}
\label{sec:data:calib2}

\begin{figure}[h]
          \begin{center}
                      \includegraphics[scale=0.6]{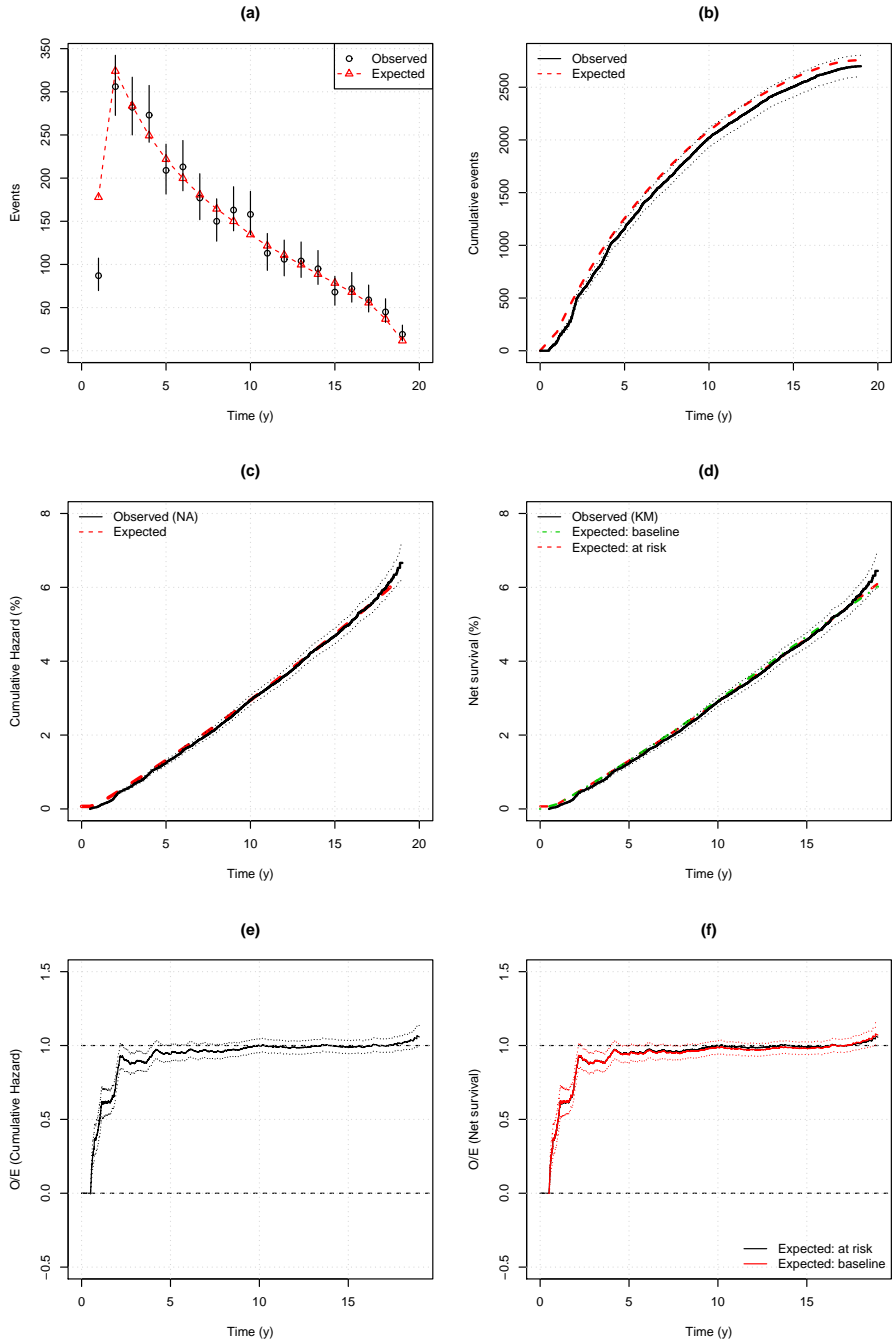}
                          \caption{Overall calibration of the breast cancer risk model. (a) Observed (95\%CI) vs expected number of breast cancers diagnosed for each year of follow-up, (b) cumulative observed (95\%CI) vs expected number of breast cancers diagnosed; (c) observed (Nelson-Aalan, 95\%CI) vs expected cumulative hazards; (d) observed (Kaplan-Meier, 95\%CI) vs expected (obtained via two methods) net risks; (e) Observed divided by Expected cumulative hazard (95\%CI); (f) Observed divided by expected net risk (obtained via two methods) with 95\%CI only for the expected risk based on baseline risk assessment. }
                                    \label{fig_calib1}
                                      \end{center}
                                            \end{figure}

Time-dependent calibration was assessed using the methods in Section \ref{sec:tdcalib}. We present results from overall model calibration (Figure \ref{fig_calib1}) and in the highest and lowest risk deciles (Supplementary Figures S1 and S2) following \cite{2018-brentnalletal}.

Overall the model appeared well calibrated throughout follow-up time. As follow-up started at six months, the expected hazard in the first year was much lower than the second year of followup. Figure \ref{fig_calib1}a shows the model did not adequately take into account the effect of removal of a pool of cancers diagnosed at entry. However, subsequently the model tracked the observed number quite closely. Figures \ref{fig_calib1}c,d show respectively the Nelson-Aalan and Kaplan-Meier curves. The corresponding observed to expected plots in Figures \ref{fig_calib1}e,f are virtually identical using the expected risk based on those still at risk. There is also very little difference between the two methods to obtain expected risk using the Kaplan-Meier approach, which indirectly confirms that censoring was unrelated to risk assessment.

The same plots were used to assess high and low risk groups. Supplementary Figure S1 considers women who were in the highest predicted risk decile at entry. As noted in Section \ref{sec:data:calib1}, there is some evidence of over-estimation from the model for this group. However, the observed to expected plots in Supplementary Figure S1e,f show a fairly constant pattern across time, with no discernible difference between Nelson-Aalan or Kaplan-Meier comparisons. Supplementary Figure S2 shows calibration of the bottom decile, where calibration appears reasonable through follow up time, albeit with a suggestion that the model had a tendency to under estimate risks.

\subsection{Regression analysis}
The above analysis suggests that the risk model was calibrated overall in the cohort, but tended to overestimate risk in those predicted at highest risk; it over-estimated risk in the first year due to omitting screen-detected cases, but the risk predictions thereafter were stable with greater follow-up time. All these aspects may be jointly tested through a Poisson regression model. 
The model was fitted with an exponential link function. The log hazard per year (or until the event or censoring if in that year) was included as an offset term, and calibration coefficients were estimated for (i) overall lack of calibration in the first year, (ii) stability of calibration from year two onward (follow-up year), and (iii) 10y risk categories from the model as previously defined ($<$2\%, 2 to $<$3\%, 3 to $<5\%$, 5 to $<8\%$, $\geq 8\%$; with 2-3\% group as the reference category) \cite{2018-brentnalletal}. 

Results are shown in Table \ref{tbl:calib}, taking the exponent of the estimated model parameter to give a calibration coefficient where unity indicates ideal calibration. They confirm the earlier analysis. The risk model over estimated risk in the first year by a factor of two (calibration coefficient 0.50, 95\%CI 0.40-0.63), which is due to the expected short-term reduction in risk following a negative screen. There was no loss in calibration subsequently (calibration 1.00, 95\%CI 0.99-1.01). The highest risk category (10y risk $\geq 8$\%) exhibited evidence of over-estimation of risk relative to the reference average-risk category (calibration overall 0.80, 95\%CI 0.69 - 0.92, \textit{i.e.} including the intercept), and likewise there was some under-estimation in the lowest-risk group (overall test for calibration across the five groups likelihood-ratio $\chi^2=48.6$, df=4). 

\section{Conclusion}

In this paper we reviewed the statistical foundations of a model to evaluate breast cancer risk at different follow-up times. This model, and corresponding computer program, synthesise findings from many scientific studies by combining risk factors into a hybrid genetic segregation and regression model. The model has been used to guide entry criteria into prevention trials \cite{2015_Cuzicketal}, and is recommended by several organisations to guide preventive and early detection strategies, including the American Cancer Society to help determine eligibility for supplemental breast magnetic resonance imaging \cite{2007_ACSMRI}. 

We developed new graphical methods for looking at calibration over follow-up time that can be combined with Poisson regression analysis. The impact of using the different methods to assess the calibration of cumulative absolute risk was reviewed. Different methods to account for competing risks when assessing risk model calibration have been used. We note that, given the diversity of methods in use and potential impact on findings, it is important when presenting calibration findings that a detailed description of the method for expected risk is also given \cite{2012-altmanetal}. 

There are opportunities to refine and improve the current risk model. Other factors are known to affect the risk of breast cancer, but most have some difficulties for inclusion in a relatively simple model. Using weight change from age 20 or weight at a young age in addition to current weight does appear to improve risk assessment to some extent \cite{Hidayat2018}, but requires accurate recall of previous weight. Alcohol consumption is a well documented risk factor \cite{2002-collaborativeover}, but underestimation of consumption is well known and it is not clear how best to allow for this. The differing roles of weight in the pre- and postmenopausal periods also needs further investigation \cite{Schoemaker2018}. Physical activity and dietary factors are also known to affect risk \cite{Prentice2006, Wu2013}, but identification of the key factors and simple accurate measures of these currently present difficulties. Levels of oestrogens and testosterone may also be important especially for postmenopausal women, but these require a blood sample and are not routinely available \cite{Zhang2018}. Ethnic differences in breast cancer risk are also well documented - notably in the US lower risks in Hispanics and Asians \cite{Banegas2017}. However it is unclear as to how these are related to differences in the known risk factors in the model, so would not need model recalibration, or whether there are intrinsic ethnic differences which require use of a separate baseline hazard function, and possibly separate risks for the known factors. Differences between recent migrants and established individuals are also known to exist and complicate any model adjustments \cite{John2005}. Adding breast density has substantially improved the model, and might be extended by considering longitudinal measurements of density and features in the mammogram other than simply breast density, where more complex algorithms may be able to extract more risk information \cite{Wang2017}. The polygenic (SNP) risk score is steadily being improved as more SNPs are added, but this is easily accommodated as all that is required is the relative risk. A bigger challenge is to accurately account for interactions between risk factors. Our model was developed by synthesizing risks for individual factors from large overviews of many studies, and large datasets which contain all the factors will be needed to fully explore interactions. These are not currently available but are likely to be so soon, as the risk factors used in our model are currently being collected in several large cohorts of patients. 

In conclusion, we are gradually moving towards an era of precision medicine, where disease treatment, early detection and prevention strategies will depend on risk assessment. The accuracy of risk models will underpin this approach. 

\section*{Acknowledgements}

This study was supported by grant C569/A16891 from Cancer Research UK. We thank Diana Buist and Erin Bowles for substantial contributions to acquisition and interpretation of data. Our use in this paper of data from Kaiser Permanente Washington Health Research Institute was supported by research specialist award R50CA211115 from the National Cancer Institute (NCI) (Ms Bowles); grants HHSN261201100031C and P01CA154292 from the NCI-funded Breast Cancer Surveillance Consortium (Dr Buist); and contracts N01-CN-005230, N01-CN-67009, N01-PC-35142, HHSN261201000029C, and HHSN261201300012I from the Cancer Surveillance System of the Fred Hutchinson Cancer Research Center, funded by the Surveillance, Epidemiology and End Results Program of the NCI with additional support from the Fred Hutchinson Cancer Research Center and the State of Washington.

\bibliography{mybib}

\begin{table}
      \centering
      \caption{Summary of risk factor parameters in model. \label{ref-tbl1}}
   \begin{tabular}{ccrrr}
\hline
     &Category&Hazard ratio&Mean risk&Reference category\\
       \hline
       \multicolumn{2}{l}{(a) Classic risk factors}\\
      \hspace{1em} Menopause age (y)& per 5y &1.14&1.08&45-49y\\
      &&&&\\
       \hspace{1em} Menarche age (y)&$<$11&1.16&1&13y\\
       &11&1.07&&\\
       &12&1.07&&\\
       &13&1&&\\
       &14&0.98&&\\
       &15&0.93&&\\
       &16&0.88&&\\
       &17 or older&0.81&&\\
       \hspace{1em}        Height (m)&$<$1.6&1&1.1&$<$1.6m\\
        &1.6-1.7&1.05+2*(height-1.6)&&\\
        &1.7 or taller&1.24&&\\
       \hspace{1em}         Body mass index&$<$21&1&1.24&$<$21\\ 
                (kg/m$^2$) &21 to $<$23&1.14&&\\
        (post-menopausal only)  &23 to $<$25&1.15&&\\
         &25 to $<$27&1.26&&\\
         &27 or more&1.32&&\\
       \hspace{1em}           Age at 1st &Nulliparous&1&1&Nulliparous\\
       childbirth (y)    &$<$17-19&0.74&&\\
           &20-24&0.77&&\\
           &25-29&0.87&&\\
           &30-34&1.01&&\\
           &35+&1.11&&\\
       \hspace{1em}            Menopausal &Not current&1&1&Not current\\
       hormone therapy             &Estrogen-only (current)&1.4&&\\
            &Combined (current)&2&\\
            &&&&\\
       \hspace{1em}             Benign disease&Non-proliferative / none&1&1&None\\
             &Proliferative (usual type)&2&&\\
             &Atypical hyperplasia&4&&\\
             &Lobular carcinoma \textit{in situ}&8&&\\
       \multicolumn{2}{l}{(b) New in v8}\\
       \hspace{1em}              Breast density residual &Visual asssessment scale&~1.4 per SD &1&Average density\\
       \hspace{1em}              (age 40y+)&BI-RADS density&~1.4 per SD && (age and BMI \\
              &Volumetric percentage&~1.4 per SD && adjusted)\\
              &&&&\\
       \hspace{1em}               SNPs&Continuous&Input&1&Average woman\\
\hline
       \multicolumn{5}{l}{Footnote: SD, standard deviation; SNP, single nucleotide polymorphism risk score.}
   \end{tabular}
 \end{table}

 \begin{table}
      \centering
      \caption{Calibration assessment: univariate and Poisson regression (adjusted for variables listed) calibration coefficient estimates with 95\% Wald confidence intervals.} \label{tbl:calib}
       \begin{tabular}{lrrrrr}
              \hline
               Term &n& O & E & O/E [univariate] & O/E (95\%CI) [adjusted] \\
                  \hline
                   Overall (intercept) & 132139 & 2699 & 2757 & 0.98 & 1.05 (0.94-1.16) \\
 Follow-up time \\
           { }                      Year 1 & 132139 & 87 & 178 & 0.49 & 0.50 (0.40-0.63) \\
           { }                         Year 2+ (time) & 123830 & 2612 & 2579 & 1.01 & 1.00 (0.99-1.01) \\
                             10y Risk group \\

                                { }$<$2\% &53436 &  641 & 548 & 1.12 & 1.17 (1.00-1.25) \\
                          { } 2 to $<$3\% & 33269&627  & 603 & 1.04 & 1  \\
                                { }  3 to $<$5\% & 29477 & 779 & 784 & 0.99 & 0.96 (0.86-1.07) \\
                                  { }   5 to $<$8\%& 11312 & 379 & 473 & 0.80 & 0.78 (0.68-0.88) \\
                                    { }    8\%+ & 4645 & 273 & 349 & 0.78 & 0.76 (0.66-0.88) \\
                                            \hline
                                             \end{tabular}
                                              \end{table}

\end{document}


\begin{frontmatter}

\title{Supplementary material for `Risk models for breast cancer and their validation'}
\runtitle{Breast cancer risk models supplement}


    \author{\fnms{Adam} R \snm{Brentnall}\ead[label=e1]{a.brentnall@qmul.ac.uk}},
    \author{\fnms{Jack} \snm{Cuzick}\corref{}\ead[label=e2]{j.cuzick@qmul.ac.uk}}

\address{Centre for Cancer Prevention, Wolfson Institute of Preventive Medicine, Queen Mary University of London, Charerhouse square, London, EC1M 6BQ \printead{e1,e2}.}

    \runauthor{Brentnall \& Cuzick}

\end{frontmatter}
\begin{figure}[h]
          \begin{center}
                      \includegraphics[scale=0.6]{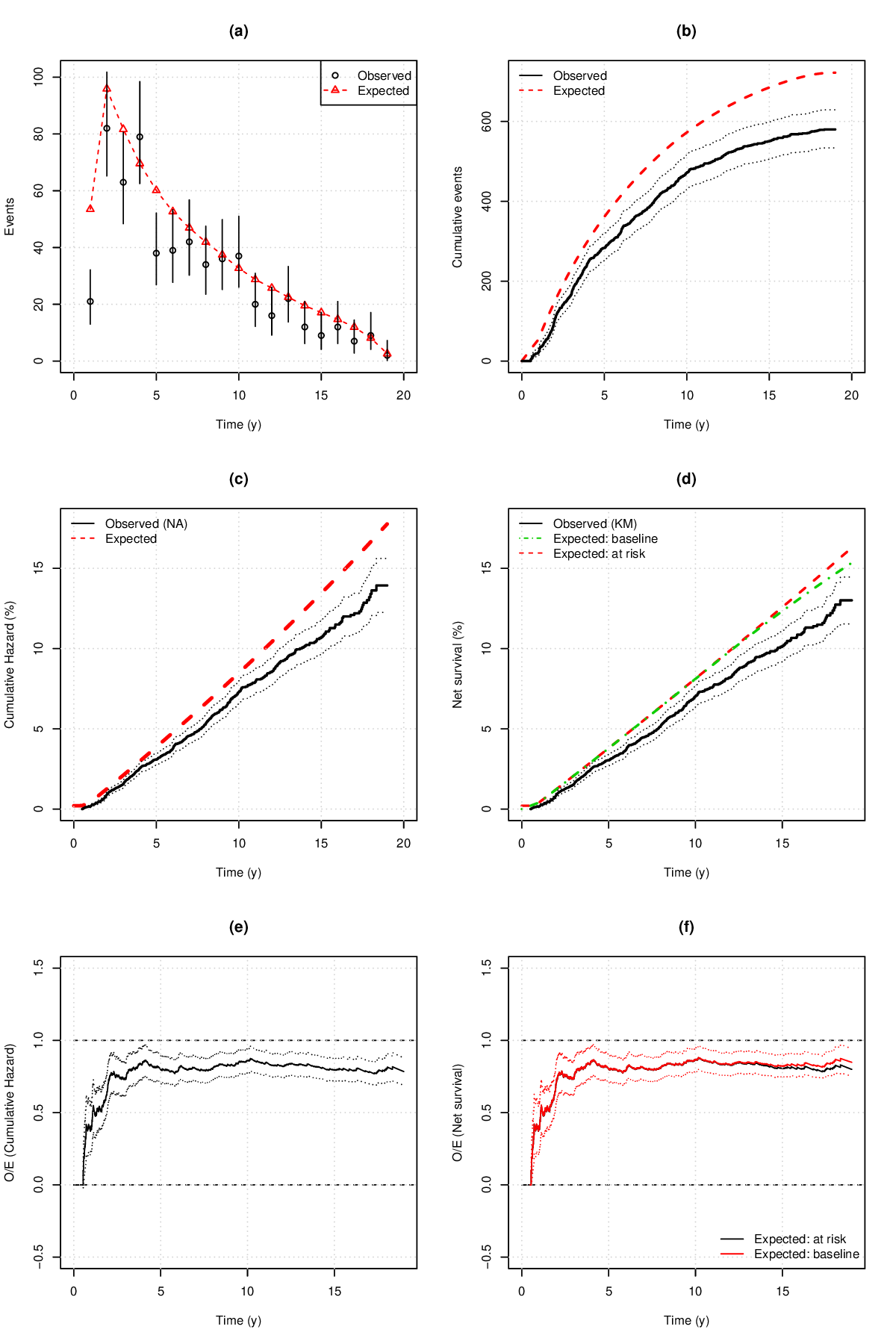}
                          \caption{Calibration of top decile of 10y predicted risk ($>$5.4\% 10y net risk). (a) Observed (95\%CI) vs expected number of breast cancers diagnosed for each year of follow-up, (b) cumulative observed (95\%CI) vs expected number of breast cancers diagnosed; (c) observed (Nelson-Aalan, 95\%CI) vs expected cumulative hazards; (d) observed (Kaplan-Meier, 95\%CI) vs expected (obtained via two methods) net risks; (e) Observed divided by Expected cumulative hazard (95\%CI); (f) Observed divided by expected net risk (obtained via two methods) with 95\%CI only for the expected risk based on baseline risk assessment. }

                                    \label{fig_calib2}
                                      \end{center}
                                            \end{figure}

%

\begin{figure}[h]
          \begin{center}
                      \includegraphics[scale=0.6]{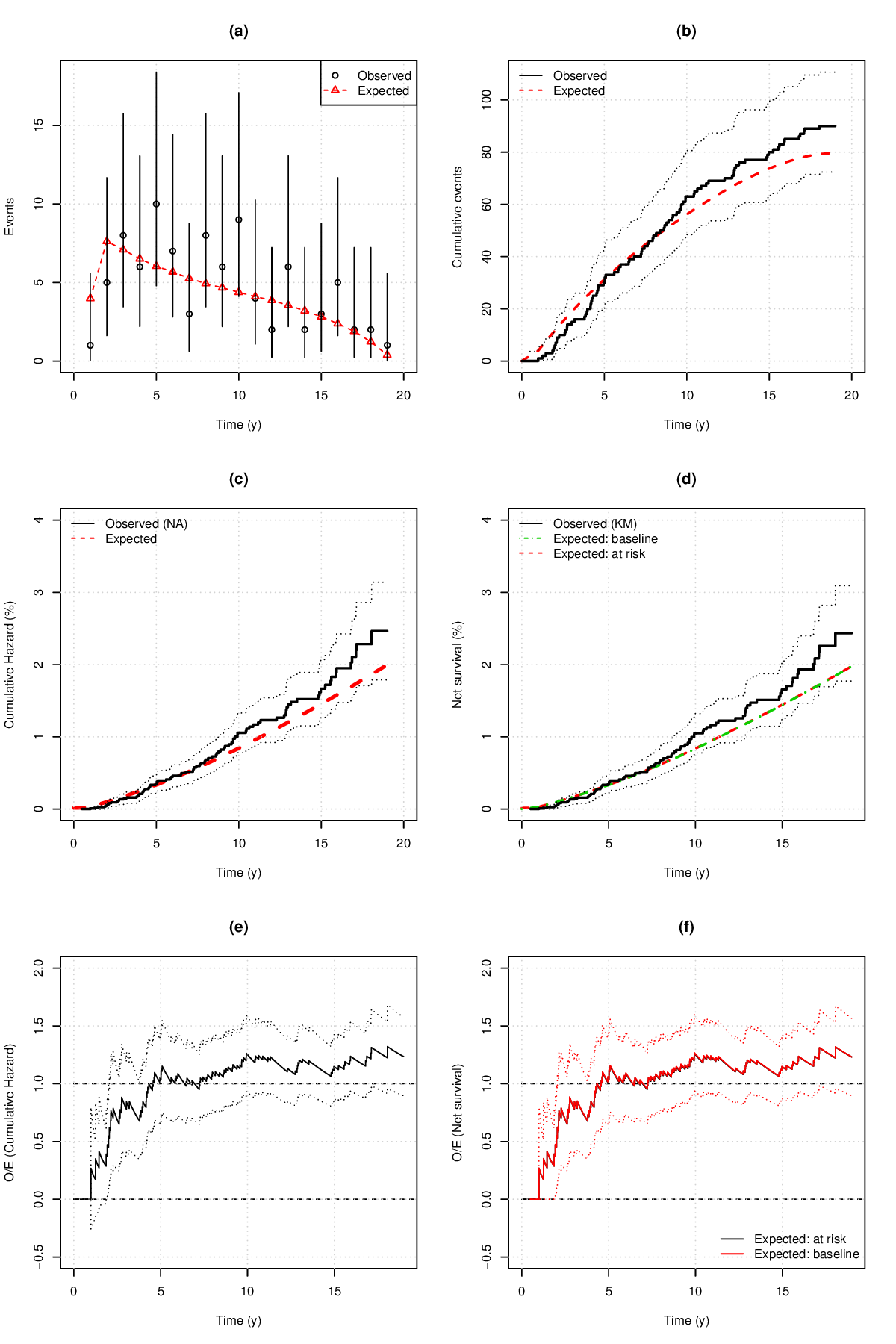}
                           \caption{Calibration of bottom decile of 10y predicted risk ($<$1.1\% 10y net risk). (a) Observed (95\%CI) vs expected number of breast cancers diagnosed for each year of follow-up, (b) cumulative observed (95\%CI) vs expected number of breast cancers diagnosed; (c) observed (Nelson-Aalan, 95\%CI) vs expected cumulative hazards; (d) observed (Kaplan-Meier, 95\%CI) vs expected (obtained via two methods) net risks; (e) Observed divided by Expected cumulative hazard (95\%CI); (f) Observed divided by expected net risk (obtained via two methods) with 95\%CI only for the expected risk based on baseline risk assessment. }
                                    \label{fig_calib4}
                                      \end{center}
                                            \end{figure}